# Dynamic Speed Harmonization

Ozgenur Kavas-Torris, Levent Guvenc
Automated Driving Lab, Ohio State University

## Abstract

In the last decade, the accelerated advancements in manufacturing techniques and material science enabled the automotive industry to manufacture commercial vehicles at more affordable rates. This, however, brought about roadways having to accommodate an ever-increasing number of vehicles every day. However, some roadways, during specific hours of the day, had already been on the brink of reaching their capacity to withstand the number of vehicles travelling on them. Hence, overcrowded roadways create slow traffic, and sometimes, bottlenecks. In this paper, a Dynamic Speed Harmonization (DSH) algorithm that regulates the speed of a vehicle to prevent it from being affected by bottlenecks has been presented. First, co-simulations were run between MATLAB Simulink and CarSim to test different deceleration profiles. Then, Hardware-in-the-Loop (HIL) simulations were run with a Road Side Unit (RSU), which emulated a roadside detector that spotted bottlenecks and sent information to the Connected Vehicle about the position of the queue and the average speed of the vehicles at the queue. The DSH algorithm was also tested on a track to compare the performance of the different deceleration profiles in terms of ride comfort.

## Introduction

In the last decade, the accelerated advancements in manufacturing techniques and material science enabled the automotive industry to manufacture commercial vehicles at more affordable rates. This, however, brought about roadways having to accommodate an ever-increasing number of vehicles every day. However, some roadways, during specific hours of the day, had already been on the brink of reaching their capacity to withstand the number of vehicles travelling on them. Hence, overcrowded roadways create slow traffic, and sometimes, bottlenecks.

Dynamic Speed Harmonization (DSH) can be utilized to overcome the mobility problems, improve roadway safety and ride comfort for passengers. DSH algorithms are able to regulate the speed of a vehicle to prevent it from being affected by bottlenecks [1], [2], [3]. Using DSH, vehicles can approach bottlenecks and queues with smoother deceleration profiles with improved drive and ride comfort.

Unmanned Aerial Vehicles (UAVs) are part of Intelligent Transportation Systems (ITSs) and can be used in a wide variety of tasks. UAVs can be deployed in several transportation related areas including traffic monitoring, traffic control, incident monitoring, transportation of goods and weather monitoring [4], [5], [6]. UAVs that are equipped with automotive grade communication devices, such as DSRC modems and 4G internet, are called Connected Unmanned Aerial Vehicles (C-UAVs) [7]. C-UAVs can gather spatio-temporal information about queue formation and related vehicle speeds using cameras and DSRC. Using this data, C-UAVs can transmit traffic information or speed advisories to CAVs that are approaching the queues. Using the data from the C-UAVs, coordination can be established between CAVs and C-UAVs [8]. One use case scenario of a CAV and C-UAV coordination can be through the utilization of DSH to improve overall ride comfort and to avoid bottleneck congestion [9].

Vehicle connectivity can be used to receive information about the upcoming road conditions to make informed and smooth deceleration decisions by the CAVs to save fuel and help roadway mobility with DSH. Unmanned Aerial Vehicles (UAVs) can be utilized for ground vehicle and aerial vehicle connectivity. Kavas-Torris *et al.* designed a system with ground and aerial vehicle coordination, where information about a roadway emergency was broadcast by the UAV and picked up by the CAV [9].

In this paper, a Dynamic Speed Harmonization scenario was prepared, where the queue information, such as the distance to queue, average vehicle speed around the queue and the total length of the queue were broadcast by either a Roadside Unit (RSU) or another On-board Unit (OBU) acting like an RSU. Three different models were prepared, where the control algorithm calculates different deceleration profiles using the information received about the traffic conditions around the queue. Step, Step-Sigmoid and Single-Sigmoid are the 3 models that generate speed trajectories for the ego CAV using the upcoming queue information received through V2I or V2X. the performance of the system was tested through Model-in-the-Loop (MIL) simulation, Hardware-in-the-Loop (HIL) simulations, Vehicle-in-the-Loop (VIL) stationary and track testing.

The remainder of the paper is organized as follows: Methodology section shows details about the DSH models developed for this work. Simulation Results section shows the performance of each DSH algorithm. Conclusions and Future Work section summarizes the work done, draws conclusions about the study and elaborates on how to extend this study for future work.

## Methodology

For the analysis, 3 different DSH models called Step, Step-Sigmoid and Single-Sigmoid were designed.

### Step DSH Algorithm

In the DSH scenario, once the ego vehicle is within the communication range of the OBU, the instantaneous vehicle speed is recorded and called $v_{DSH\,start}$ (1). Next, the speed difference between the $v_{DSH\,start}$ and queue speed ($v_{queue}$) is calculated (2). The roadway is then divided into equ-length segments (3).

$$v_{DSH\,start} = v_{instantaneous} \text{ when } d_{queue\,start} - d_{travelled} < d_{OBU\,range} \quad (1)$$

$$v_{difference} = v_{DSH\ start} - v_{queue} \quad (2)$$

$$d_{segment} = floor\left(\frac{d_{queue\ end} - d_{queue\ start}}{\frac{v_{difference}}{a_{deceleration}}}\right) \quad (3)$$

Depending on how many segments there are between the current location of the vehicle and the queue location, while also taking the desired deceleration limit into account, a recommended speed profile is calculated for each segment.

For the Step DSH algorithm speed profile, firstly the number of segments is calculated (4). Then, the vehicle decelerates at the amount of the desired deceleration limit at each section (5). When the Step recommended speed is lower than or equal to the queue speed acquired from V2I or V2X, then the recommended speed is equal to the queue speed (6). Once the queue is passed, then the number of segments to reach back up to the speed limit is calculated (7). Then, the vehicle accelerates at the amount of the desired acceleration limit at each section (8). When the Step recommended speed is higher than or equal to the speed limit, then the recommended speed is set to be equal to the speed limit (9).

$$count_{dec} = ceil\left(\frac{d_{travelled} - d_{queue\ start}}{d_{segment}}\right) * (d_{travelled} - d_{queue\ start}) \quad (4)$$

$$v_{Step} = v_{vehicle} - a_{deceleration} * count_{dec} \quad (5)$$

$$v_{Step} = v_{queue}\ when\ v_{Step} \leq v_{queue} \quad (6)$$

$$count_{acc} = ceil\left(\frac{d_{travelled} - d_{queue\ end}}{d_{segment}}\right) * (d_{travelled} - d_{queue\ end}) \quad (7)$$

$$v_{Step} = v_{vehicle} + a_{acceleration} * count_{acc} \quad (8)$$

$$v_{Step} = v_{speed\ limit}\ when\ v_{Step} \geq v_{speed\ limit} \quad (9)$$

### Step-Sigmoid DSH Algorithm

For the Step-Sigmoid DSH algorithm speed profile, the number of road segments to reach the queue speed is calculated the same way in Equation (4). Then, the incremental sigmoid deceleration trend is calculated (10) and the recommended speed for each segment is found for when the vehicle is decelerating to approach the queue (11). Similarly, the incremental sigmoid acceleration trend is calculated (12) when the vehicle is accelerating and leaving the queue location (13).

$$v_{inc_{dec}} = \frac{1}{1 + e^{-0.09*\left(d_{travelled} - \left(d_{OBU\ range} + \left((count_{dec}+1)*d_{segment} + \frac{d_{seg}}{}\right)\right)\right)}} \quad (10)$$

$$v_{Step-Sigmoid} = v_{vehicle} - a_{deceleration} * v_{inc_{dec}}\ when\ decelerating \quad (11)$$

$$v_{inc_{acc}} = \frac{1}{1 + e^{-0.09*\left(d_{travelled} - \left(d_{OBU\ range} + \left((count_{acc}+1)*d_{segment} + \frac{d_{seg}}{}\right)\right)\right)}} \quad (12)$$

$$v_{Step-Sigmoid} = v_{vehicle} + a_{acceleration} * v_{inc_{acc}}\ when\ accelerating \quad (13)$$

### Single-Sigmoid DSH Algorithm

For the Single-Sigmoid DSH algorithm speed profile, when the ego vehicle is approaching the queue, the Single-Sigmoid recommended incremental deceleration profile is calculated depending on the speed limit, desired deceleration rate, queue speed, start of the queue and end of queue location (14). The Single-Sigmoid deceleration profile is found using equation (15). Similarly, the incremental acceleration profile is calculated (16) and the Single-Sigmoid acceleration profile is found (17).

$$v_{inc_{dec}} = \frac{1}{1 + e^{-0.009*\left(d_{travelled} - \left(d_{queue\ start} + \left(d_{queue\ end} + \frac{d_{queue\ start}}{2}\right)\right)\right)}} \quad (14)$$

$$v_{Single-Sigmoid} = v_{vehicle} - v_{difference} * v_{inc_{dec}}\ when\ decelerating \quad (15)$$

$$v_{inc_{acc}} = \frac{1}{1 + e^{-0.009*\left(d_{travelled} - \left(d_{queue\ end} + \left(d_{queue\ end} + \frac{d_{queue\ start}}{2}\right)\right)\right)}} \quad (16)$$

$$v_{Single-Sigmoid} = v_{vehicle} + v_{difference} * v_{inc_{acc}}\ when\ accelerating \quad (17)$$

# Simulation Results

The 3 different DSH algorithms were modelled in Simulink. For the performance evaluation of the DSH algorithm, Model-in-the-Loop (MIL) co-simulations were run firstly between MATLAB Simulink and CarSim to test different deceleration profiles. Then, Hardware-in-the-Loop (HIL) simulations were run with a Road Side Unit (RSU), which emulated a roadside detector that spotted bottlenecks and sent information to the Connected Vehicle (CV) about the position of the queue and the average speed of the traffic vehicles at the queue. The DSH algorithms were also tested on an actual track with the CAV development platform vehicle to see how it affected ride comfort for passengers.

## *Model-in-the-Loop (MIL) Results*

The DSH algorithms with 3 different profiles, namely Step, Step-Sigmoid and Single-Sigmoid, were tested in a CarSim – Simulink MIL environment. The ego vehicle controls with DSH were run on Simulink and CarSim was used for realistic high fidelity vehicle dynamics of the test vehicle.

The results of the DSH MIL simulation can be seen in Figure 1. During the simulation, after the test vehicle travelled for 4,200 m, the vehicle received information about an upcoming queue 1,000 m ahead. In the 1$^{st}$ sub-plot, it is seen that DSH was not active, and the vehicle decelerated rapidly once the driver realized that there was a queue ahead. In the 2$^{nd}$ subplot, the Step DSH algorithm commanded the vehicle to decelerate to the queue speed of 5 m/s, drive at constant speed in the queue and accelerate with the Step speed profile. In the 3$^{rd}$ subplot, the Single-Sigmoid DSH commanded the vehicle to decelerate to the queue speed of 5 m/s, drive at constant speed in the queue and accelerate with the Single Sigmoid speed profile. In the 4$^{th}$ subplot, the Step-Sigmoid DSH commanded the vehicle to decelerate to the queue speed of 5 m/s, drive at constant speed in the queue and accelerate with the Step Sigmoid speed profile.

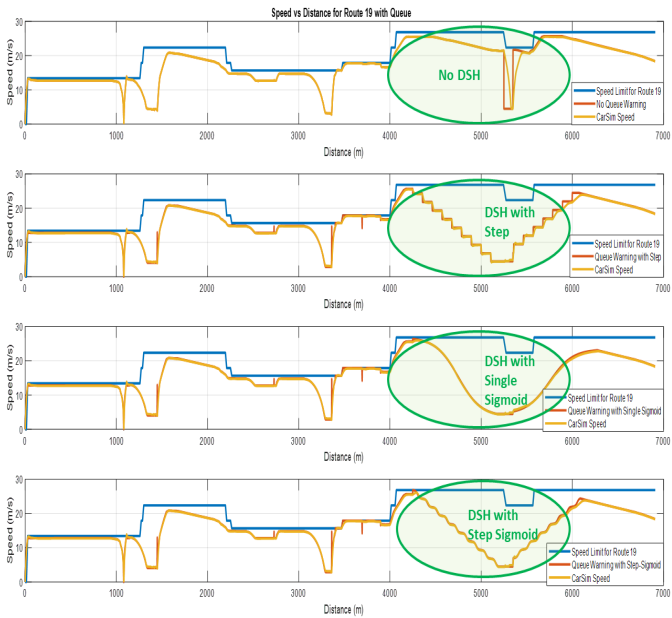

*Figure 1. Dynamic Speed Harmonization in MIL results*

Looking at the results presented in Figure 1, it is seen that using DSH enabled both smoother acceleration and deceleration compared to the case of no DSH (1$^{st}$ subplot). During the no DSH case, the vehicle decelerated rapidly when it reached the queue and could not decelerate smoothly. The fuel consumption values were similar to each other for all 4 cases. However, the drive and ride comfort was better for the 3 cases where the DSH was active.

## *Hardware-in-the-Loop (HIL) Results*

The HIL environment was set up to test the DSH model (Figure 2). For the HIL test, two DSRC modems were used. The first modem acted as a Roadside Unit (RSU), where the RSU gets information about the queue location and the average vehicle speed at the queue. The second modem was used to act as the ego CAV, which received information about the queue from the RSU. The MABX received the queue information from the ego vehicle OBU and used it with the DSH algorithms running on it. Then, ego vehicle control commands were sent to the SCALEXIO real time system acting as the actual vehicle, and realistic vehicle and simulation data was sent back to MABX to update DSH controls.

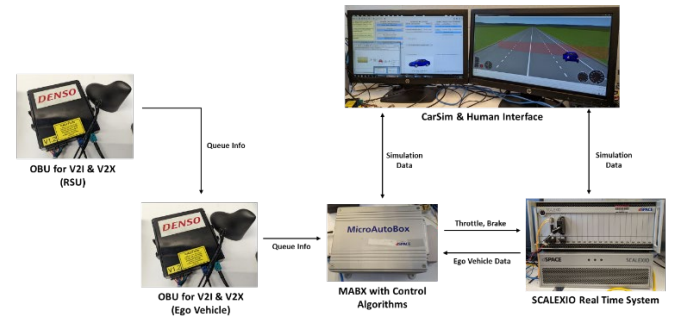

*Figure 2. HIL setup for DSH testing with V2I & V2X*

The results of the DSH in HIL simulation can be seen below in Figure 3. The no DSH case seen in the 1$^{st}$ subplot demanded the driver to decelerate very rapidly to the queue speed, which was an aggressive maneuver, and the vehicle could not decelerate rapidly enough to follow the desired speed profile. Using Step DSH and Single-Sigmoid DSH allowed the ego vehicle to decelerate down to the desired queue speed smoothly. Moreover, the ego vehicle was able to follow the desired speed more closely.

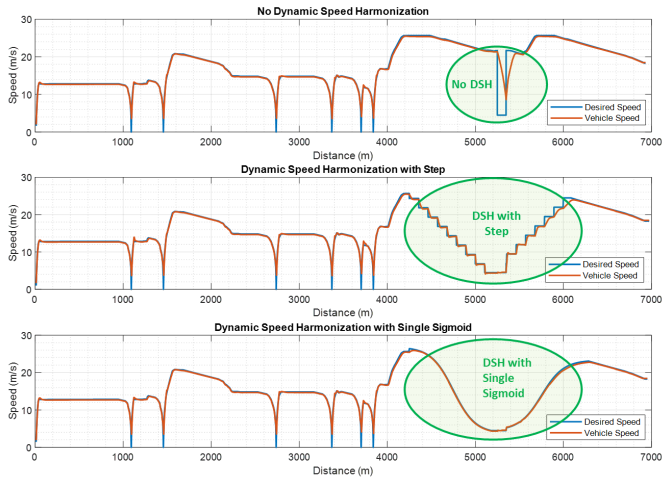

*Figure 3. Dynamic Speed Harmonization in HIL results*

### Vehicle-in-the-Loop (VIL) Stationary Testing Results

This VIL implementation was similar to the HIL implementation and schematic of this VIL environment can be seen in Figure 4. Rather than using the SCALEXIO real time system to act as the actual vehicle and CarSim for human interface, the actual ego vehicle was used. Connections to the CAN buses of the vehicle through CAN setup blocks allowed access to actual vehicle data. dSPACE ControlDesk interface was utilized to visualize and record the test data. The OBU mounted on the CAV received queue information through DSRC from the OBU attached to the UAV. The Step (advisory speed trajectory in yellow), Step-Sigmoid (advisory speed trajectory in green) and Single-Sigmoid (advisory speed trajectory in red) DSH algorithms running on the MABX used the queue information to calculate smooth speed, as seen in Figure 4.

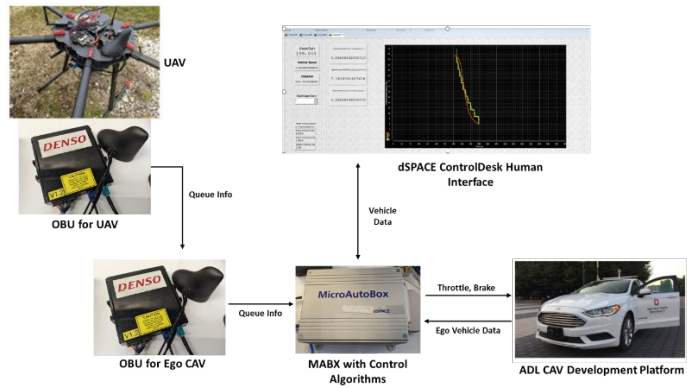

*Figure 4. VIL setup for DSH testing with V2I & V2X*

### Vehicle-in-the-Loop (VIL) Track Testing Results

Vehicle testing was conducted on the 8-mile round track at the Transportation Research Center (TRC) using the CAV development platform Ford Fusion hybrid vehicle of ADL.

The results of the VIL track test can be seen in Figure 5. During the testing, the vehicle was given a reference speed profile to follow (the blue line). The actual speed of the vehicle is seen as the red line. The speed limit of the soft test route can be seen as the black dashed line in the plot.

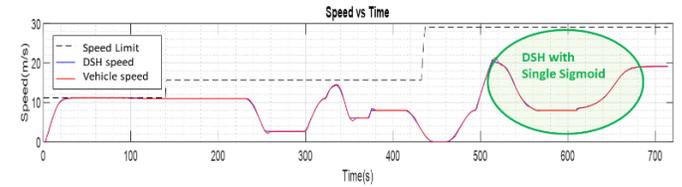

*Figure 5. Dynamic Speed Harmonization in VIL testing on track*

As seen in Figure 5, after the vehicle travelled for 510 seconds, the DSH algorithm was activated. Once the DSH was activated, the vehicle received information that there was going to be a bottleneck ahead. To guarantee smooth deceleration, drive and ride comfort, the vehicle was commanded by the Single-Sigmoid DSH algorithm to decelerate from 20 m/s to 10 m/s with the single sigmoid deceleration profile. When the vehicle reached the queue location, it was commanded to travel at constant speed. Then, around 610 seconds, the ego vehicle passed the queue and the bottleneck, and accelerated up to 20 m/s again with the single sigmoid acceleration profile. During the test, all 4 passengers in the ADL CAV 2017 Fusion vehicle were satisfied with the smooth deceleration and acceleration profile, when the DSH with Single Sigmoid was active.

## Conclusions and Future Work

In this paper, 3 different DSH algorithms were developed for mobility and safety improvement on roadways. The DSH algorithms were tested with MIL and HIL simulations and VIL track testing to generate an advisory speed profile for the ego vehicle approaching a queue. Secondly, a communication link between a CAV and UAV was established to show how CAV – UAV coordination can be utilized as part of an Intelligent Transportation System (ITS). The DSH algorithm was also tested as part of the CAV – UAV connected system with VIL tests.

Looking at the results, it is shown that transportation systems can be more connected and more efficient through ground and aerial vehicle coordination. Using DSH, smooth advisory trajectories can be recommended for drivers to follow, so that queues and bottlenecks are prevented from forming and disrupting the traffic flow, and mobility of roadway vehicles are improved.

For future work, the DSH formulation can be improved to get communication delay into account. Tracking of the desired DSH speed profile while avoiding collisions with unexpected obstacles or while platooning, if needed, can be implemented and improved by using advanced and robust controls and connected and autonomous driving related methods [10-62]

## Contact Information

Ozgenur Kavas-Torris, Ph.D., is a member of the Automated Driving Lab (ADL, https://mekar.osu.edu/), Department of Mechanical and Aerospace Engineering (MAE), College of Engineering, The Ohio State University, Columbus, OH 43210 USA (e-mail: kavastorris.1@osu.edu).

# References

[1] K. Balke, H. Charara, and S. Sunkari, "Report on dynamic speed harmonization and queue warning algorithm design.," no. FHWA-JPO-14-168, Feb. 2014, [Online]. Available: https://rosap.ntl.bts.gov/view/dot/3534

[2] A. T. Hamdar Hani S. Mahmassani, Samer H., "Speed Harmonization: Evaluation of Effectiveness Under Congested Conditions - Alireza Talebpour, Hani S. Mahmassani, Samer H. Hamdar, 2013," Transportation Research Record, Jan. 2013, Accessed: Jun. 29, 2020. [Online]. Available: https://journals-sagepub-com.proxy.lib.ohio-state.edu/doi/10.3141/2391-07

[3] H. Mahmassani, H. Rakha, E. Hubbard, D. Lukasik, and Science Applications International Corporation, "Concept development and needs identification for Intelligent Network Flow Optimization (INFLO) : assessment of relevant prior and ongoing research.," FHWA-JPO-13-011, Mar. 2012. Accessed: Nov. 17, 2021. [Online]. Available: https://rosap.ntl.bts.gov/view/dot/3380

[4] J. Cecil, "A conceptual framework for supporting UAV based cyber physical weather monitoring activities," in 2018 Annual IEEE International Systems Conference (SysCon), Apr. 2018, pp. 1–8. doi: 10.1109/SYSCON.2018.8369588.

[5] P. Cheng, G. Zhou, and Z. Zheng, "Detecting and Counting Vehicles from Small Low-Cost UAV Images," 2009.

[6] H. Menouar, I. Guvenc, K. Akkaya, A. S. Uluagac, A. Kadri, and A. Tuncer, "UAV-Enabled Intelligent Transportation Systems for the Smart City: Applications and Challenges," IEEE Communications Magazine, vol. 55, no. 3, pp. 22–28, Mar. 2017, doi: 10.1109/MCOM.2017.1600238CM.

[7] G. Dowd, O. Kavas-Torris, L. Guvenc, and B. Aksun-Guvenc, "Simulation Environment for Visualizing Connected Ground and Air Traffic - Garrett Dowd, Ozgenur Kavas-Torris, Levent Guvenc, Bilin Aksun-Guvenc, 2020," Transportation Research Record, Apr. 2021, Accessed: Oct. 23, 2020. [Online]. Available: http://journals.sagepub.com/doi/10.1177/0361198120962485

[8] O. Kavas-Torris, S. Y. Gelbal, M. R. Cantas, B. A. Guvenc, and L. Guvenc, "Connected UAV and CAV Coordination for Improved Road Network Safety and Mobility," presented at the SAE WCX Digital Summit 2021, Warrendale, PA, Apr. 2021. Accessed: Mar. 31, 2021. [Online]. Available: https://www.sae.org/publications/technical-papers/content/2021-01-0173/

[9] Kavas-Torris, O., Gelbal, S.Y., Cantas, M.R., Aksun-Guvenc, B., Guvenc, L., "V2X Communication Between Connected and Automated Vehicles (CAVs) and Unmanned Aerial Vehicles (UAVs)," Sensors, 22, 8941. https://doi.org/10.3390/s22228941.

[10] S. Y. Gelbal, M. R. Cantas, B. A. Guvenc, L. Guvenc, G. Surnilla, H. Zhang, M. Shulman, A. Katriniok and J. Parikh, "Hardware-in-the-Loop and Road Testing of RLVW and GLOSA Connected Vehicle Applications," in SAE WCX, 2020.

[11] S. Gelbal, B. Aksun-Guvenc and Guvenc, L., "Elastic Band Collision Avoidance of Low Speed Autonomous Shuttles with Pedestrians," International Journal of Automotive Technology, vol. 21, no. 4, pp. 903-917, 2020.

[12] S. Gelbal, S. Arslan, H. Wang, B. Aksun-Guvenc and L. Guvenc, "Elastic Band Based Pedestrian Collision Avoidance using V2X Communication," in IEEE Intelligent Vehicles Symposium, Redondo Beach, California, 2017.

[13] H. Wang, A. Tota, B. Aksun-Guvenc and L. Guvenc, "Real time implementation of socially acceptable collision avoidance of a low speed autonomous shuttle using the elastic band method," Mechatronics, vol. 50, pp. 341-355, 2018.

[14] L. Guvenc, B. Aksun-Guvenc, S. Zhu and S. Gelbal, Autonomous Road Vehicle Path Planning and Tracking Control, New York: Wiley / IEEE Press, Book Series on Control Systems Theory and Application, 2002.

[15] H. Wang, S. Gelbal and L. Guvenc, ""Multi-Objective Digital PID Controller Design in Parameter Space and its Application to Automated Path Following," IEEE Access, vol. 9, pp. 46874-46885, 2021.

[16] B. Demirel and L. Guvenc, "Parameter Space Design of Repetitive Controllers for Satisfying a Mixed Sensitivity Performance Requirement," IEEE Transactions on Automatic Control, vol. 55, pp. 1893-1899, 2010.

[17] B. Aksun-Guvenc and L. Guvenc, "Robust Steer-by-wire Control based on the Model Regulator," in IEEE Conference on Control Applications, 2002.

[18] B. Orun, S. Necipoglu, C. Basdogan and L. Guvenc, "State Feedback Control for Adjusting the Dynamic Behavior of a Piezo-actuated Bimorph AFM Probe," Review of Scientific Instruments, vol. 80, no. 6, 2009.

[19] L. Guvenc and K. Srinivasan, "Friction Compensation and Evaluation for a Force Control Application," Journal of Mechanical Systems and Signal Processing, vol. 8, no. 6, pp. 623-638.

[20] M. Emekli and B. Aksun-Guvenc, "Explicit MIMO Model Predictive Boost Pressure Control of a Two-Stage Turbocharged Diesel Engine," IEEE Transactions on Control Systems Technology, vol. 25, no. 2, pp. 521-534, 2016.

[21] Li, X., Arul Doss, A.C., Aksun-Guvenc, B., Guvenc, L., 2020, "Pre-Deployment Testing of Low Speed, Urban Road Autonomous Driving in a Simulated Environment," SAE International Journal of Advances and Current Practices in Mobility.

[22] Guvenc, L., Aksun-Guvenc, B., Li, X., Arul Doss, A.C., Meneses-Cime, K.M., Gelbal, S.Y., 2019, Simulation Environment for Safety Assessment of CEAV Deployment in Linden, Final Research Report, Smart Columbus Demonstration Program – Smart City Challenge Project (to support Contract No. DTFH6116H00013).

[23] Kavas-Torris, O., Cantas, M., Meneses Cime, K., Aksun Guvenc, B., Guvenc L., "The Effects of Varying Penetration Rates of L4-L5 Autonomous Vehicles on Fuel Efficiency and Mobility of Traffic Networks," SAE Technical Paper 2020-01-0137, 2020, https://doi.org/10.4271/2020-01-0137.

[24] Cantas, M., Fan, S., Kavas, O., Tamilarasan, S., Guvenc, L., Yoo, S., Lee J., Lee, B., Ha, J. "Development of Virtual Fuel Economy Trend Evaluation Process," SAE Technical Paper 2019-01-0510, 2019, https://doi.org/10.4271/2019-01-0510.

[25] Kavas-Torris, O., Lackey, N.A., Guvenc, L., "Simulating Autonomous Vehicles in a Microscopic Traffic Simulator to Investigate the Effects of Autonomous Vehicles on Roadway Mobility," International Journal of Automotive Technology, in press.

[26] Emirler, M.T., Uygan, I.M.C., Aksun Guvenc, B., Guvenc, L., 2014, "Robust PID Steering Control in Parameter Space for Highly Automated Driving," International Journal of Vehicular Technology, Vol. 2014, Article ID 259465.

[27] Emirler, M.T., Wang, H., Aksun Guvenc, B., Guvenc, L., 2015, "Automated Robust Path Following Control based on Calculation of Lateral Deviation and Yaw Angle Error," ASME Dynamic Systems and Control Conference, DSC 2015, October 28-30, Columbus, Ohio, U.S.

[28] Zhou, H., Jia, F., Jing, H., Liu, Z., Guvenc, L., 2018, "Coordinated Longitudinal and Lateral Motion Control for

Four Wheel Independent Motor-Drive Electric Vehicle," IEEE Transactions on Vehicular Technology, Vol. 67, No 5, pp. 3782-379.
[29] Emirler, M.T., Kahraman, K., Senturk, M., Aksun Guvenc, B., Guvenc, L., Efendioglu, B., 2015, "Two Different Approaches for Lateral Stability of Fully Electric Vehicles," International Journal of Automotive Technology, Vol. 16, Issue 2, pp. 317-328.
[30] S Zhu, B Aksun-Guvenc, Trajectory Planning of Autonomous Vehicles Based on Parameterized Control Optimization in Dynamic on-Road Environments, Journal of Intelligent & Robotic Systems, 1-13, 2020.
[31] Gelbal, S.Y., Cantas, M.R, Tamilarasan, S., Guvenc, L., Aksun-Guvenc, B., 2017, "A Connected and Autonomous Vehicle Hardware-in-the-Loop Simulator for Developing Automated Driving Algorithms," IEEE Systems, Man and Cybernetics Conference, Banff, Canada.
[32] Boyali A., Guvenc, L., 2010, "Real-Time Controller Design for a Parallel Hybrid Electric Vehicle Using Neuro-Dynamic Programming Method," IEEE Systems, Man and Cybernetics, İstanbul, October 10-13, pp. 4318-4324.
[33] Kavas-Torris, O., Cantas, M.R., Gelbal, S.Y., Aksun-Guvenc, B., Guvenc, L., "Fuel Economy Benefit Analysis of Pass-at-Green (PaG) V2I Application on Urban Routes with STOP Signs," Special Issue on Safety and Standards for CAV, International Journal of Vehicle Design, in press.
[34] Yang, Y., Ma, F., Wang, J., Zhu, S., Gelbal, S.Y., Kavas-Torris, O., Aksun-Guvenc, B., Guvenc, L., 2020, "Cooperative Ecological Cruising Using Hierarchical Control Strategy with Optimal Sustainable Performance for Connected Automated Vehicles on Varying Road Conditions," Journal of Cleaner Production, Vol. 275, in press.
[35] Hartavi, A.E., Uygan, I.M.C., Guvenc, L., 2016, "A Hybrid Electric Vehicle Hardware-in-the-Loop Simulator as a Development Platform for Energy Management Algorithms," International Journal of Vehicle Design, Vol. 71, No. 1/2/3/4, pp. 410-420.
[36] Emirler, M.T., Kahraman, K., Senturk, M., Aksun Guvenc, B., Guvenc, L., Efendioğlu, B., 2013, "Estimation of Vehicle Yaw Rate Using a Virtual Sensor," International Journal of Vehicular Technology, Vol. 2013, ArticleID 582691.
[37] Kavas-Torris, O., Gelbal, S.Y., Cantas, M.R., Aksun-Guvenc, B., Guvenc, L., "V2X Communication Between Connected and Automated Vehicles (CAVs) and Unmanned Aerial Vehicles (UAVs)," Sensors, 22, 8941. https://doi.org/10.3390/s22228941.
[38] Meneses-Cime, K., Aksun-Guvenc, B., Guvenc, L., 2022, "Optimization of On-Demand Shared Autonomous Vehicle Deployments Utilizing Reinforcement Learning," Sensors, 22, 8317. https://doi.org/10.3390/s22218317.
[39] Wang, J., Wu, G., Sun, B., Ma, F., Aksun-Guvenc, B., Guvenc, L., 2022, "Disturbance Observer-Smith Predictor Compensation based Platoon Control with Estimation Deviation," Journal of Advanced Transportation, Vol. 2022, Article ID 9866794, 14 pages, https://doi.org/10.1155/2022/9866794.
[40] Ma, F., Yang, Y., Wang, J., Li, X., Wu, G., Zhao, Y., Wu, L., Aksun-Guvenc, B., Guvenc, L., 2021, "Eco-Driving-Based Cooperative Adaptive Cruise Control of Connected Vehicles Platoon at Signalized Intersections," Transportation Research Part D: Transport and Environment, Vol. 92, 102746, ISSN 1361-9209, https://doi.org/10.1016/j.trd.2021.102746.
[41] Wang, J., Ma, F., Yu, Y., Nie, J., Aksun-Guvenc, B., Guvenc, L., 2020, "Adaptive Event Triggered Platoon Control under Unreliable Communication Links," IEEE Transactions on Intelligent Transportation Systems, doi: 10.1109/TITS.2020.3030016.
[42] Dowd, G.E., Kavas-Torris, O., Guvenc, L., Aksun-Guvenc, B., 2020, "Simulation Environment for Visualizing Connected Ground and Air Traffic," Transportation Research Record: Journal of the Transportation Research Board, October, doi:10.1177/0361198120962485.
[43] Ma, F., Wang, J., Yang Y., Liang W., Zhenze L., Aksun-Guvenc, B., Guvenc, L., 2020, "Parameter-space-based Robust Control of Event-triggered Heterogeneous Platoon," IET Intelligent Transport Systems, Vol. 15, pp. 61–73, https://doi.org/10.1049/itr2.12004.
[44] Kavas-Torris, O., Cantas, M.R., Gelbal, S.Y., Aksun-Guvenc, B., Guvenc, L., 2020, "Fuel Economy Benefit Analysis of Pass-at-Green (PaG) V2I Application on Urban Routes with STOP Signs," Special Issue on Safety and Standards for CAV, International Journal of Vehicle Design, Vol. 83, No. 2/3/4, pp. 258-279.
[45] Wang, J., Ma, F., Yu, Y., Zhu, S., Gelbal, S.Y., Aksun-Guvenc, B., Guvenc, L., 2020, "Optimization Design of the Decentralized Multi-Vehicle Cooperative Controller for Freeway Ramp Entrance," International Journal of Automotive Technology, Vol. 22, pp. 799–810.
[46] Li, X., Arul Doss, A.C., Aksun-Guvenc, B., Guvenc, L., 2020, "Pre-Deployment Testing of Low Speed, Urban Road Autonomous Driving in a Simulated Environment," SAE International Journal of Advances and Current Practices in Mobility, Vol. 2, Issue 6, pp. 3301-3311, doi.org/10.4271/2020-01-0706.
[47] Yang, Y., Ma, F., Wang, J., Zhu, S., Gelbal, S.Y., Kavas-Torris, O., Aksun-Guvenc, B., Guvenc, L., 2020, "Cooperative Ecological Cruising Using Hierarchical Control Strategy with Optimal Sustainable Performance for Connected Automated Vehicles on Varying Road Conditions," Journal of Cleaner Production, Vol. 275, doi.org/10.1016/j.jclepro.2020.123056.
[48] Ma, F., Wang, J., Zhu, S., Gelbal, S.Y., Yu, Y., Aksun-Guvenc, B., Guvenc, L., 2020, "Distributed Control of Cooperative Vehicular Platoon with Nonideal Communication Condition," IEEE Transactions on Vehicular Technology, Vol. 69, Issue 8, pp. 8207-8220, doi:10.1109/TVT.2020.2997767.
[49] Ding, Y., Zhuang, W., Wang, L., Liu, J., Guvenc, L., Li, Z., 2020, "Safe and Optimal Lane Change Path Planning for Automated Driving," IMECHE Part D Passenger Vehicles, Vol. 235, No. 4, pp. 1070-1083, doi.org/10.1177/0954407020913735.
[50] Emirler, M.T., Gözü, M., Uygan, I.M.C., Böke, T.A., Aksun-Guvenc, B., Guvenc, L., 2018, "Evaluation of Electronic Stability Controllers Using Hardware-in-the-Loop Vehicle Simulator," International Journal of Advances in Automotive Engineering, pp. 123-141.
[51] Emirler, M.T., Guvenc, L., Aksun-Guvenc, B., 2018, "Design and Evaluation of Robust Cooperative Adaptive Cruise Control Systems in Parameter Space," International Journal of Automotive Technology, Vol. 19, Issue 2, pp. 359-367.
[52] Emirler, M.T., Uygan, İ.M.C., Gelbal, Ş.Y., Gözü, M., Böke, T.A., Aksun Guvenc, B., Guvenc, L., 2016, "Vehicle Dynamics Modelling and Validation for a Hardware-in-the-Loop Vehicle Simulator," International Journal of Vehicle Design, Vol. 71, No. 1/2/3/4, pp. 191-211.
[53] Emirler, M.T., Kahraman, K., Şentürk, M., Aksun Guvenc, B., Guvenc, L., Efendioğlu, B., 2015, "Two Different Approaches for Lateral Stability of Fully Electric Vehicles," International Journal of Automotive Technology, Vol. 16, Issue 2, pp. 317-328.

[54] Necipoglu, S., Cebeci, S.A., Basdogan, Ç., Has, Y.E., Guvenc, L., 2011, "Repetitive Control of an XYZ Piezo-stage for Faster Nano-scanning: Numerical Simulations and Experiment," Mechatronics, Vol. 21, No. 6, pp. 1098-1107.

[55] Necipoglu, S., Cebeci, S.A., Has, Y.E., Guvenc, L., Basdogan, Ç., 2011, "A Robust Repetitive Controller for Fast AFM Imaging," IEEE Transactions on Nanotechnology, Vol. 10, No. 5, pp. 1074-1082.

[56] Demirel, B., Guvenc, L., 2010, "Parameter Space Design of Repetitive Controllers for Satisfying a Mixed Sensitivity Performance Requirement," IEEE Transactions on Automatic Control, Vol. 55, No. 8, pp. 1893-1899.

[57] Aksun Guvenc, B., Guvenc, L., Karaman, S., 2010, "Robust MIMO Disturbance Observer Analysis and Design with Application to Active Car Steering," International Journal of Robust and Nonlinear Control, Vol. 20, pp. 873-891.

[58] Orun, B., Necipoglu, S., Basdogan, Ç., Guvenc, L., 2009, "State Feedback Control for Adjusting the Dynamic Behavior of a Piezo-actuated Bimorph AFM Probe," Review of Scientific Instruments, Vol. 80, No. 6.

[59] Aksun Guvenc, B., Guvenc, L., Karaman, S., 2009, "Robust Yaw Stability Controller Design and Hardware in the Loop Testing for a Road Vehicle," IEEE Transactions on Vehicular Technology, Vol. 58, No. 2, pp. 555-571.

[60] Guvenc, L., Srinivasan, K., 1995, "Force Controller Design and Evaluation for Robot Assisted Die and Mold Polishing," Journal of Mechanical Systems and Signal Processing, Vol. 9, No. 1, pp. 31-49.

[61] Guvenc, L., Srinivasan, K., 1994, "Friction Compensation and Evaluation for a Force Control Application," Journal of Mechanical Systems and Signal Processing, Vol. 8, No. 6, pp. 623-638.

[62] Cebi, A., Guvenc, L., Demirci, M., Kaplan Karadeniz, C., Kanar, K., Guraslan, E., 2005, "A Low Cost, Portable Engine ECU Hardware-In-The-Loop Test System," Mini Track of Automotive Control, IEEE International Symposium on Industrial Electronics Conference, Dubrovnik, June 20-23.